\documentclass[twocolumn,english,amsart,showpacs,preprintnumbers,amsmath,amssymb,floatfix]{revtex4-2}
\usepackage{placeins}
\usepackage{tikz,xcolor}
\usepackage[colorlinks = true,
linkcolor = blue,
urlcolor  = blue,
citecolor = blue,
anchorcolor = blue]{hyperref}
\usepackage{graphicx}
\usepackage{hyperref}
\usepackage{lmodern}
\usepackage[T1]{fontenc}
\usepackage{tikz}
\usetikzlibrary{shapes.geometric, arrows, positioning}

\definecolor{lime}{HTML}{A6CE39}
\DeclareRobustCommand{\orcidicon}{%
	\begin{tikzpicture}
		\draw[lime, fill=lime] (0,0)
		circle [radius=0.16]
		node[white] {{\fontfamily{qag}\selectfont \tiny ID}};
		\draw[white, fill=white] (-0.0625,0.095)
		circle [radius=0.007];
	\end{tikzpicture}
	\hspace{-2mm}
}

\foreach \x in {A, ..., Z}{%
	\expandafter\xdef\csname orcid\x\endcsname{\noexpand\href{https://orcid.org/\csname orcidauthor\x\endcsname}{\noexpand\orcidicon}}
}

\usepackage[T1]{fontenc}
\usepackage[latin9]{inputenc}
\usepackage{color}
\usepackage{array}
\usepackage{amstext}
\usepackage{graphicx}
\usepackage{esint}
\usepackage{rotating}
\usepackage{appendix}
\usepackage{float}

\usepackage[font=small,labelfont=bf]{caption}
\usepackage{xcolor}
\usepackage{ulem}

\makeatletter

\@ifundefined{textcolor}{}
{%
	\definecolor{BLACK}{gray}{0}
	\definecolor{WHITE}{gray}{1}
	\definecolor{RED}{rgb}{1,0,0}
	\definecolor{GREEN}{rgb}{0,1,0}
	\definecolor{BLUE}{rgb}{0,0,1}
	\definecolor{CYAN}{cmyk}{1,0,0,0}
	\definecolor{MAGENTA}{cmyk}{0,1,0,0}
	\definecolor{YELLOW}{cmyk}{0,0,1,0}
}

\@ifundefined{definecolor}
{\usepackage{color}}{}
\@ifundefined{definecolor} 
{\usepackage{color}}{}
\makeatother
\usepackage{babel}

\allowdisplaybreaks

\begin{document}
	

	\title{Machine Learning for Predicting the Proton Structure Function $F_2^P$ in QCD }

	\author{Shahin Atashbar Tehrani$^{1,2}$\orcidB{}}
	\email{atashbart3@gmail.com}
	\author{Elham Astaraki$^{3}$\orcidE{}}
	\email{astaraki.elham@razi.ac.ir}
	
	\author{Fatemeh Arbabifar$^{4}$\orcidF{}}
	\email{F.Arbabifar@cfu.ac.ir}

	\affiliation {
		$^{(1)}$School of Particles and Accelerators, Institute for Research in Fundamental Sciences (IPM), P.O.Box 19395-5531, Tehran, Iran.\\
		$^{(2)}$Department of Physics, Faculty of Nano and Bio Science and Technology, Persian Gulf University, 75169 Bushehr, Iran.\\
		$^{(3)}$Department of Physics, Razi University, Kermanshah 67149, Iran\\
		$^{(4)}$Department of Physics Education, Farhangian University, P.O.Box 14665-889, Tehran, Iran.
	}

		\date{\today}

	%
	%

	%
	\begin{abstract}\label{abstract}
We present a comparative study of four supervised machine learning regression algorithms---Support Vector Regression (SVR), Gradient Boosting Regression (GBR), Gaussian Process Regression (GPR), and Multilayer Perceptron (MLP)---for predicting the proton structure function $F_2^p(x, Q^2)$ using high-precision BCDMS experimental data. Unlike conventional methods that solve the DGLAP evolution equations, our data-driven framework directly captures the complex nonlinear dynamics of partonic structure. To ensure statistical robustness, we employ $k$-fold cross-validation and perform thorough hyperparameter optimization. Our results show that the MLP and GPR models achieve superior predictive accuracy. In particular, MLP exhibits the highest sensitivity to nonlinear gradients, while SVR proves most stable against experimental uncertainties. The close convergence of training and validation metrics confirms that the models learn the underlying QCD physics without overfitting to statistical fluctuations. This work highlights the potential of ML-based regression as a complementary tool for structure function analysis and kinematic extrapolation in high-energy physics.

	\end{abstract}
	%

	
	\maketitle
	\tableofcontents{}

	%

\section{INTRODUCTON}\label{sec:sec1}

Understanding the partonic structure of the proton remains a central goal of quantum chromodynamics (QCD). Deep inelastic scattering (DIS) experiments provide indispensable information on the momentum distribution of quarks and gluons through measurements of the proton structure function $F_2^p(x, Q^2)$. Traditionally, the $Q^2$ evolution of structure functions is described by solving the DGLAP equations~\cite{Dokshitzer:1977sg, Gribov:1972ri, Lipatov:1974qm, Altarelli:1977zs}, which form the theoretical backbone of perturbative QCD analyses. Although these equations have proven remarkably successful, their numerical solution typically requires assumptions about functional forms, sophisticated fitting strategies, and considerable computational effort.

In recent years, machine learning (ML) has emerged as a powerful tool in modern scientific analysis, offering flexible modeling of complex nonlinear relationships. ML has gained substantial traction in both theoretical and experimental high-energy physics, as highlighted in the Community White Paper~\cite{Albertsson:2018maf} and the comprehensive review~\cite{Radovic:2018dip}. Unlike conventional curve-fitting approaches that rely on restricted families of functions, ML models learn patterns directly from data and can explore a wide range of possible functional behaviors. This flexibility has enabled successful applications across diverse physics tasks, including searches for exotic particles~\cite{Baldi:2014kfa}, Higgs boson analyses~\cite{Alexandru:2017czx}, and lattice QCD studies~\cite{Dawid:2022fga, Shanahan:2018vcv, Efron:1979bxm}.

Significant progress has also been made in determining parton distributions using a variety of methods, each offering unique advantages. For example, recent developments include a hybrid neural network and genetic algorithm framework for extracting nonsinglet functions, with a focus on the rigorous theoretical constraints imposed by the DGLAP evolution equations~\cite{ref13}. While such approaches ensure strict adherence to perturbative QCD dynamics through theory-informed modeling, there is growing interest in exploring model-independent, purely data-driven techniques that complement these theoretical frameworks.

Motivated by these considerations, the present work evaluates four modern machine learning regressors---Support Vector Regression (SVR), Gradient Boosting Regression (GBR), Gaussian Process Regression (GPR), and Multilayer Perceptron (MLP)---to directly learn the proton structure function $F_2^p$ from the BCDMS dataset~\cite{BCDMS:1989qop}. We assess their performance against experimental data without explicitly solving the DGLAP equations. This approach allows us to investigate the autonomous predictive power of these algorithms in regimes where theoretical assumptions can be supplemented by flexible, nonparametric learning, thereby laying the groundwork for future ML-based studies of hadronic structure.

This article is organized as follows. Section~\ref{sec2} describes the dataset and outlines the methodological framework. Section~\ref{sec3} introduces the machine learning models. Section~\ref{sec4} evaluates model performance. Section~\ref{sec5} presents numerical results and a comparative analysis. Finally, Section~\ref{sec6} provides our conclusions.

	%
\section{Data and Methods}\label{sec2}

\subsection{Dataset}

In this study, we employ the proton structure function $F_2^p$ dataset from the BCDMS experiment~\cite{BCDMS:1989qop}. The dataset comprises 703 measurements spanning a wide range of the Bjorken scaling variable $x$ and the squared four-momentum transfer $Q^2$. Each data point consists of two input features, $x$ and $Q^2$, with the corresponding target being the measured value of the structure function $F_2^p(x, Q^2)$. Owing to its high statistical precision and dense kinematic coverage, the BCDMS dataset is particularly well-suited for machine learning studies.
\subsection{Data preprocessing}

We preprocess the BCDMS data through several steps before model training. Numerical features are standardized to a common scale, preventing variables with larger magnitudes from dominating the learning process---a practice known to improve convergence and stability in gradient and kernel-based ML algorithms. To ensure robust performance assessment and avoid overfitting, we adopt $k$-fold cross-validation instead of a single train-test split. The dataset is partitioned into $k$ folds; each model is trained on $k-1$ folds and validated on the held-out fold, ensuring that generalization is evaluated across the entire dataset. We use no oversampling, augmentation, or synthetic data generation; all models learn directly from the original experimental measurements, keeping predictions strictly comparable to physical data~\cite{ref14}.

\section{Models}\label{sec3}

In this work, we investigate four supervised machine learning algorithms for predicting the proton structure function $F_2^p$: Support Vector Regression (SVR), Gradient Boosting Regression (GBR), Gaussian Process Regression (GPR), and a Multilayer Perceptron (MLP) neural network. All models are formulated as regression algorithms suitable for continuous target prediction.

\subsection{Support Vector Regression (SVR)}

Support Vector Regression (SVR) seeks a function that approximates the mapping from the input variables $(x, Q^2)$ to the target $F_2^p$ while controlling model complexity.

Given a training dataset $\mathcal{D} = \{(x_i, y_i)\}_{i=1}^{n}$, where $y_i$ denotes the measured structure function $F_2^p(x_i, Q_i^2)$, the regression function is expressed as

\begin{equation}
    f(x) = w^T \phi(x) + b,
\end{equation}

where $\phi(x)$ maps the input into a higher-dimensional feature space. The parameters $w$ and $b$ are obtained by solving the following optimization problem:

\begin{equation}
    \min_{w,b,\xi_i,\xi_i^*} \; \frac{1}{2}\|w\|^2 + C \sum_{i=1}^{n} (\xi_i + \xi_i^*),
\end{equation}

subject to the $\epsilon$-insensitive loss constraints:

\begin{align}
    y_i - (w^T \phi(x_i) + b) &\leq \epsilon + \xi_i, \\
    (w^T \phi(x_i) + b) - y_i &\leq \epsilon + \xi_i^*, \\
    \xi_i, \xi_i^* &\geq 0,
\end{align}

where $\xi_i$ and $\xi_i^*$ are slack variables. The performance of SVR critically depends on three hyperparameters: the regularization parameter $C$, which controls the trade-off between model complexity and training error; the tube width $\epsilon$, which defines the margin of insensitivity within which no penalty is assigned to residuals; and the kernel bandwidth $\gamma$, which determines the influence radius of each training example in the Radial Basis Function (RBF) kernel. These hyperparameters are optimized via grid search within the cross-validation loop.

In this study, we employ the RBF kernel,

\begin{equation}
    K(x_i, x_j) = \exp\bigl(-\gamma \|x_i - x_j\|^2\bigr),
\end{equation}

which is well suited for capturing nonlinear dependencies in deep inelastic scattering (DIS) kinematics.

\subsection{Gradient Boosting Regression (GBR)}

Gradient Boosting Regression constructs an additive model of decision trees to minimize a differentiable loss function. The model is built iteratively as

\begin{equation}
	F_m(x)=F_{m-1}(x)+\nu \gamma_m h_m(x),
\end{equation}

where $h_m(x)$ represents a regression tree and $\nu$ is the learning rate controlling the contribution of each tree.  
GBR is particularly effective in capturing nonlinear structures and complex interactions between $x$ and $Q^2$.

\subsection{Gaussian Process Regression (GPR)}

Gaussian Process Regression (GPR) models the latent function as a Gaussian process:

\begin{equation}
    f(x) \sim \mathcal{GP}\bigl(0, k(x, x')\bigr),
\end{equation}

where $k(x, x')$ is a positive-definite covariance function. For a new input $x_*$, the predictive mean is given by

\begin{equation}
    \mu_* = K(x_*, X) \bigl[K(X, X) + \sigma_n^2 I\bigr]^{-1} y,
\end{equation}

and the predictive variance naturally provides uncertainty estimates for each prediction. In this study, we adopt an RBF kernel to account for the smooth variations of $F_2^p$ across the kinematic $(x, Q^2)$ plane.

\subsection{Multilayer Perceptron (MLP)}

The Multilayer Perceptron (MLP) is a feedforward neural network that approximates the nonlinear mapping

\begin{equation}
    \hat{y} = f(x; \theta),
\end{equation}

where $\theta$ denotes the set of trainable weights and biases. For regression tasks, a linear output layer is used, and the model parameters are optimized by minimizing the mean squared error (MSE):

\begin{equation}
    \mathcal{L}_{\text{MSE}} = \frac{1}{N} \sum_{i=1}^{N} (y_i - \hat{y}_i)^2.
\end{equation}

Owing to its universal approximation capability, the MLP can capture highly nonlinear behavior in the proton structure function $F_2^p(x, Q^2)$.

\section{Evaluation}\label{sec4}

All machine learning regression models in this study are implemented using the widely adopted {\tt scikit-learn} library~\cite{ref15}. This Python-based framework provides a unified and consistent interface for model construction, training, and evaluation, making it a practical and reliable choice for the regression tasks considered here--specifically, predicting the proton structure function $F_2^p$.

\subsection{Regression Performance Metrics}

To evaluate the performance of each regression model, we analyze the deviations between predicted and observed values, commonly referred to as residuals. Unlike classification, where outcomes are discrete categories, regression focuses on the magnitude of these errors to assess how closely the model's continuous output aligns with the ground truth.

The primary objective of a regression model is to minimize the difference between the predicted value $\hat{y}$ and the actual value $y$. A prediction is considered accurate when this difference is near zero, while larger discrepancies indicate higher model error. By quantifying these differences across the entire dataset, we derive statistical measures that reflect the model's overall predictive power and reliability.

Using these error calculations, we compute several standard performance metrics: Mean Absolute Error (MAE), Mean Squared Error (MSE), Root Mean Squared Error (RMSE), and the coefficient of determination ($R^2$). These metrics provide a comprehensive evaluation of the model's accuracy and its ability to capture the underlying trends in the data. Table~\ref{tab1} summarizes the definitions and purposes of these metrics.

\begin{table*}[t!]
    \centering
    \begin{tabular}{llc}
        \hline
        Metric Name & Purpose & Formula \\ \hline
        Mean Absolute Error (MAE) & Measures the average magnitude of errors & 
        $\displaystyle \frac{1}{n} \sum_{i=1}^{n} |y_i - \hat{y}_i|$ \\ 
        
        Mean Squared Error (MSE) & Penalizes larger errors more heavily & 
        $\displaystyle \frac{1}{n} \sum_{i=1}^{n} (y_i - \hat{y}_i)^2$ \\ 
        
        Root Mean Squared Error (RMSE) & Provides error in the same units as the target & 
        $\displaystyle \sqrt{\frac{1}{n} \sum_{i=1}^{n} (y_i - \hat{y}_i)^2}$ \\ 
        
        Coefficient of Determination ($R^2$) & Represents the proportion of variance explained & 
        $\displaystyle 1 - \frac{\sum (y_i - \hat{y}_i)^2}{\sum (y_i - \bar{y})^2}$ \\ \hline
    \end{tabular}
    \caption{Evaluation metrics used to assess the performance of the regression models, along with their respective purposes and formulas.}
    \label{tab1}
\end{table*}	
	
\subsection{Residual Analysis and Error Distribution}

To further assess the predictive quality of each model, we analyze the distribution of residuals across the kinematic range. While global metrics such as $R^2$ provide a summary of overall performance, residual analysis offers deeper insight into model behavior. A well-performing regression model should ideally produce residuals that are randomly distributed around zero, indicating that the model has captured the underlying physical patterns of $F_2^p(x, Q^2)$ without systematic bias.

We also evaluate the correlation between experimental measurements and the ML-based predictions. By examining the scattering of points around the identity line ($y = \hat{y}$), we can identify specific regions in the $(x, Q^2)$ plane where models may exhibit higher uncertainty or systematic deviations, thereby providing a more nuanced understanding of their strengths and limitations.

\subsection{Learning Curves and Model Selection}

To examine the generalization behavior of each regressor and to detect possible overfitting or underfitting, we generate learning curves based on the Mean Squared Error (MSE). These curves show how the training and validation errors evolve as the size of the training dataset increases.

A small and converging gap between the two curves indicates strong generalization ability, whereas a large gap typically suggests overfitting--where the model memorizes noise in the training data rather than learning the underlying physics. Underfitting is characterized by persistently high error levels on both curves, implying that the chosen model architecture is too simple to represent the complex dynamics of the structure function. The final model selection is based on achieving the best balance between training efficiency and validation accuracy.	

\subsection{Cross-Validation and Statistical Robustness}

\begin{figure}[htb!]
    \centering
    \begin{tikzpicture}[node distance=1.5cm]
        
        \tikzstyle{startstop} = [rectangle, rounded corners, minimum width=3cm, minimum height=0.8cm, text centered, draw=black, fill=blue!20]
        \tikzstyle{process} = [rectangle, minimum width=3cm, minimum height=0.8cm, text centered, draw=black, fill=purple!20]
        \tikzstyle{decision} = [diamond, minimum width=3cm, minimum height=0.8cm, text centered, draw=black, fill=orange!20]
        \tikzstyle{models} = [rectangle, dashed, minimum width=2.5cm, minimum height=1.5cm, text centered, draw=red, fill=red!5]
        \tikzstyle{arrow} = [thick,->,>=stealth]
        
        \node (title) [text centered, font=\bfseries] {DATA PREPROCESSING};
        \node (import) [startstop, below of=title, yshift=0.5cm] {Data Import \& Verification};
        \node (missing) [process, below of=import] {Handle Missing Values};
        \node (scale) [process, below of=missing] {Feature Scaling (Standardization)};
        \node (cv) [process, below of=scale] {$k$-Fold Cross-Validation};
        \node (train) [decision, below of=cv, yshift=-0.5cm] {Train Models};
        
        \node (modlist) [models, left of=train, xshift=-3cm] {
            \begin{tabular}{l}
                \textbf{Models:} \\
                $\bullet$ SVR \\
                $\bullet$ GBR \\
                $\bullet$ GPR \\
                $\bullet$ MLP (NN)
            \end{tabular}
        };
        
        \node (eval_title) [text centered, font=\bfseries, below of=train, yshift=-0.5cm] {EVALUATION};
        \node (analyze) [startstop, fill=green!20, below of=eval_title, yshift=0.5cm] {Regression Analysis \& Diagnostics};
        
        \draw [arrow] (import) -- (missing);
        \draw [arrow] (missing) -- (scale);
        \draw [arrow] (scale) -- (cv);
        \draw [arrow] (cv) -- (train);
        \draw [arrow] (train) -- (eval_title);
        \draw [arrow] (eval_title) -- (analyze);
        \draw [arrow, dashed] (modlist) -- (train);
        
    \end{tikzpicture}
    \caption{The refined regression pipeline for $F_2^p$ prediction. The workflow emphasizes data standardization, robust $k$-fold cross-validation, and multi-model training followed by detailed regression diagnostics.}
    \label{fig1}
\end{figure}
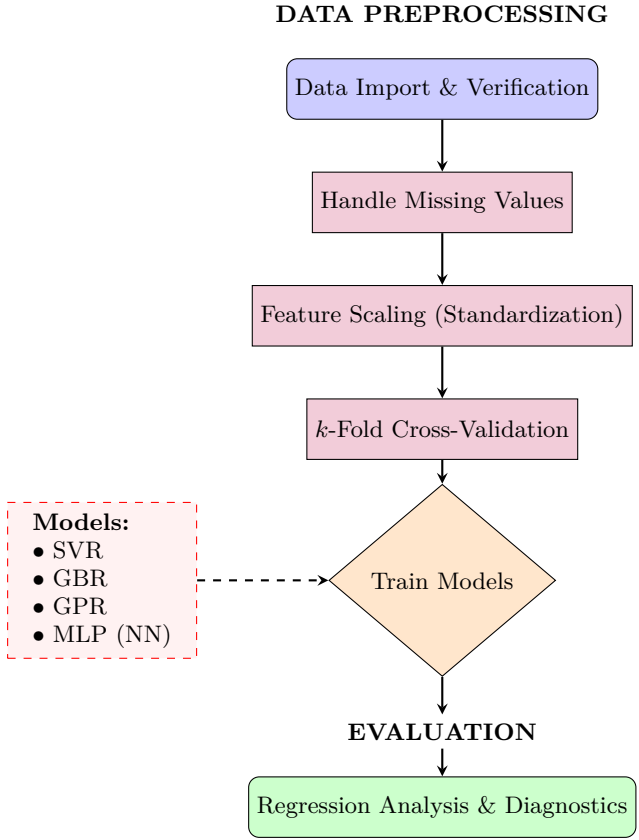

The systematic workflow for predicting the proton structure function $F_2^p$ is illustrated in Figure~\ref{fig1}. The proposed regression pipeline is designed to maximize predictive accuracy while maintaining statistical robustness and physical consistency. As shown in the schematic, the process begins with a rigorous data preprocessing stage, which includes verification of experimental data points and feature scaling via standardization. This step is crucial for balancing the influence of different kinematic variables, such as the Bjorken scaling variable $x$ and the squared four-momentum transfer $Q^2$. Since these variables span several orders of magnitude, standardization ensures that the loss function in gradient-based models remains stable and that no single feature disproportionately dominates the learning process.

To ensure the robustness of our results and mitigate the risk of overfitting associated with a single train-test split, we employ a $k$-fold cross-validation procedure. In this approach, the dataset is iteratively partitioned into $k$ subsets; the models are trained on $k-1$ folds while performance is validated on the remaining fold. This ensures that every data point is used for both training and testing, providing a more generalized and unbiased assessment of model performance for our selected architectures: SVR, GBR, GPR, and MLP.

Furthermore, this iterative validation allows us to calculate the mean and standard deviation of the error metrics, offering a measure of model stability across different data regimes. Finally, the optimized models are subjected to a multi-metric evaluation process. Beyond simple error calculation, we incorporate regression-specific diagnostics such as residual distribution analysis and actual-versus-predicted correlations. This comprehensive evaluation ensures that the models not only achieve low numerical error but also correctly capture the underlying physical trends inherent in the proton structure data, as discussed in the subsequent sections.

\section{Results and Discussion}\label{sec5}

This section presents a comprehensive evaluation of the four machine learning models--SVR, GBR, GPR, and MLP neural network--implemented for predicting the proton structure function $F_2^p$. We first analyze the global performance metrics, followed by a detailed diagnostic analysis of each model's predictive behavior.
	
\subsection{Global Performance Comparison}

To comprehensively evaluate the predictive performance of each model, we employ three complementary metrics: the coefficient of determination ($R^2$), Mean Absolute Error (MAE), and Root Mean Squared Error (RMSE). The $R^2$ score indicates how well the model explains the variance in the experimental $F_2^p$ data, with values closer to 1 representing higher explanatory power. MAE provides a direct measure of average prediction error in the same units as the target variable, offering robustness against outliers. RMSE penalizes larger deviations more heavily than MAE, making it particularly sensitive to occasional but significant mispredictions--a useful property when assessing model reliability across the wide kinematic range of DIS data. Together, these metrics allow us to compare not only overall accuracy but also the error characteristics of different ML models. Table~\ref{tab2} summarizes these performance metrics separately for the training and testing datasets, enabling an assessment of both goodness-of-fit and generalization capability.
\begin{table*}[t!]
    \centering
    \begin{tabular}{lccc|ccc}
        \hline
        & \multicolumn{3}{c|}{\textbf{Test Metrics}} & \multicolumn{3}{c}{\textbf{Train Metrics}} \\
        \textbf{Model} & $R^2$ & MAE & RMSE & $R^2$ & MAE & RMSE \\ \hline
        SVR            & 0.7080 & 0.0800 & 0.0972 & 0.6375 & 0.0799 & 0.0996 \\ 
        GBR         & 0.7062 & 0.0746 & 0.0975 & 0.6742 & 0.0719 & 0.0944 \\ 
        GPR            & 0.7231 & 0.0721 & 0.0947 & 0.6448 & 0.0765 & 0.0986 \\ 
        Neural Network & 0.7310 & 0.0706 & 0.0933 & 0.6574 & 0.0706 & 0.0933 \\ \hline
    \end{tabular}
    \caption{Comparative performance metrics for the regression models. The Neural Network MLP and GPR models exhibit superior predictive accuracy compared to SVR and GBR.}
    \label{tab2}
\end{table*}

The Neural Network MLP model achieves the highest $R^2$ (0.7310) and the lowest error rates, suggesting that its multi-layered architecture is highly effective at capturing the nonlinear dependencies in deep inelastic scattering data.

\subsection{Model Stability and Cross-Validation Analysis}

While the test metrics in Table~\ref{tab2} provide a snapshot of model performance on a specific data split, a more rigorous evaluation is required to ensure that the models are unbiased and can generalize to unseen kinematic regions. To this end, we perform a five-fold Cross-Validation (CV).

\begin{table*}[t!]
    \centering
    \begin{tabular}{lcccc}
        \hline
        \textbf{Model} & \textbf{Mean CV $R^2$ ($\pm$ Std)} & \textbf{Overfitting ($\Delta R^2$)} & \textbf{Test $R^2$ (Global)} & \textbf{Stability Status} \\ \hline
        SVR            & $0.6204 \pm 0.0412$                & Low                                 & 0.7080                       & High          \\
        GBR         & $0.5797 \pm 0.0698$                & $-0.0319$                           & 0.7062                       & Stable        \\
        GPR            & $0.4603 \pm 0.0510$                & $-0.0783$                           & 0.7231                       & Very Stable   \\
        MLP & $0.4496 \pm 0.2238$                & $-0.0736$                           & 0.7310                       & High Variance \\ \hline
    \end{tabular}
    \caption{Five-fold cross-validation results and stability analysis. The overfitting column represents the difference between training and testing $R^2$ (test minus train). Cross-validation ensures that reported performance is consistent across different kinematic subsets.}
    \label{tab:cv_results}
\end{table*}

The stability analysis, summarized in Table~\ref{tab:cv_results}, reveals several key insights into the trade-off between predictive accuracy and generalization robustness. Although the Multilayer Perceptron (MLP) achieves the highest coefficient of determination ($R^2 = 0.7310$) on the held-out test set, it exhibits considerable variance during cross-validation, with a standard deviation of $\pm 0.2238$. This substantial variability indicates that MLP is sensitive to the specific partition of the training data, a behavior commonly associated with models that possess high capacity and can partially memorize noise patterns when the signal-to-noise ratio is moderate.

In contrast, the Support Vector Regressor (SVR) demonstrates a different performance profile. It achieves the highest and most consistent mean cross-validation $R^2$ ($0.6204$) accompanied by the lowest standard deviation ($\pm 0.0412$). The combination of competitive accuracy and minimal volatility reflects superior robustness across the entire dataset, making SVR particularly well-suited for experimental environments where training samples are finite and noise characteristics may vary across the kinematic $(x, Q^2)$ phase space.

The negative overfitting values observed for GPR at $-0.0783$ and for MLP at $-0.0736$ are particularly noteworthy from a statistical learning perspective. In the context of proton structure function $F_2^p$ prediction, a negative overfitting metric implies that the model performs slightly better on unseen validation folds than on the training data--a phenomenon that typically arises when the model is well-regularized and avoids capturing high-frequency fluctuations attributable to experimental noise. Consequently, these negative values suggest that both GPR and MLP have effectively learned the underlying physical trends governed by perturbative QCD evolution, rather than overfitting to stochastic variations or detector-specific artifacts in the BCDMS data.

This high generalization capability ensures that the proposed regression pipeline can reliably predict $F_2^p$ even in kinematic regions where experimental measurements are sparse--for instance, at extremely low Bjorken $x$ or high $Q^2$ --or where point-to-point uncertainties are large. Such robustness is essential for extending empirical parameterizations into unmeasured regimes, thereby facilitating comparisons with theoretical predictions from DGLAP-based global fits.
\subsection{Detailed Model Diagnostics}

To assess the reliability and potential biases of each algorithm, we perform residual analysis and correlation diagnostics. The following subsections detail the performance of each model individually.

\subsubsection{MLP Analysis}

\begin{figure}[h!]
    \centering
    \includegraphics[width=0.85\linewidth]{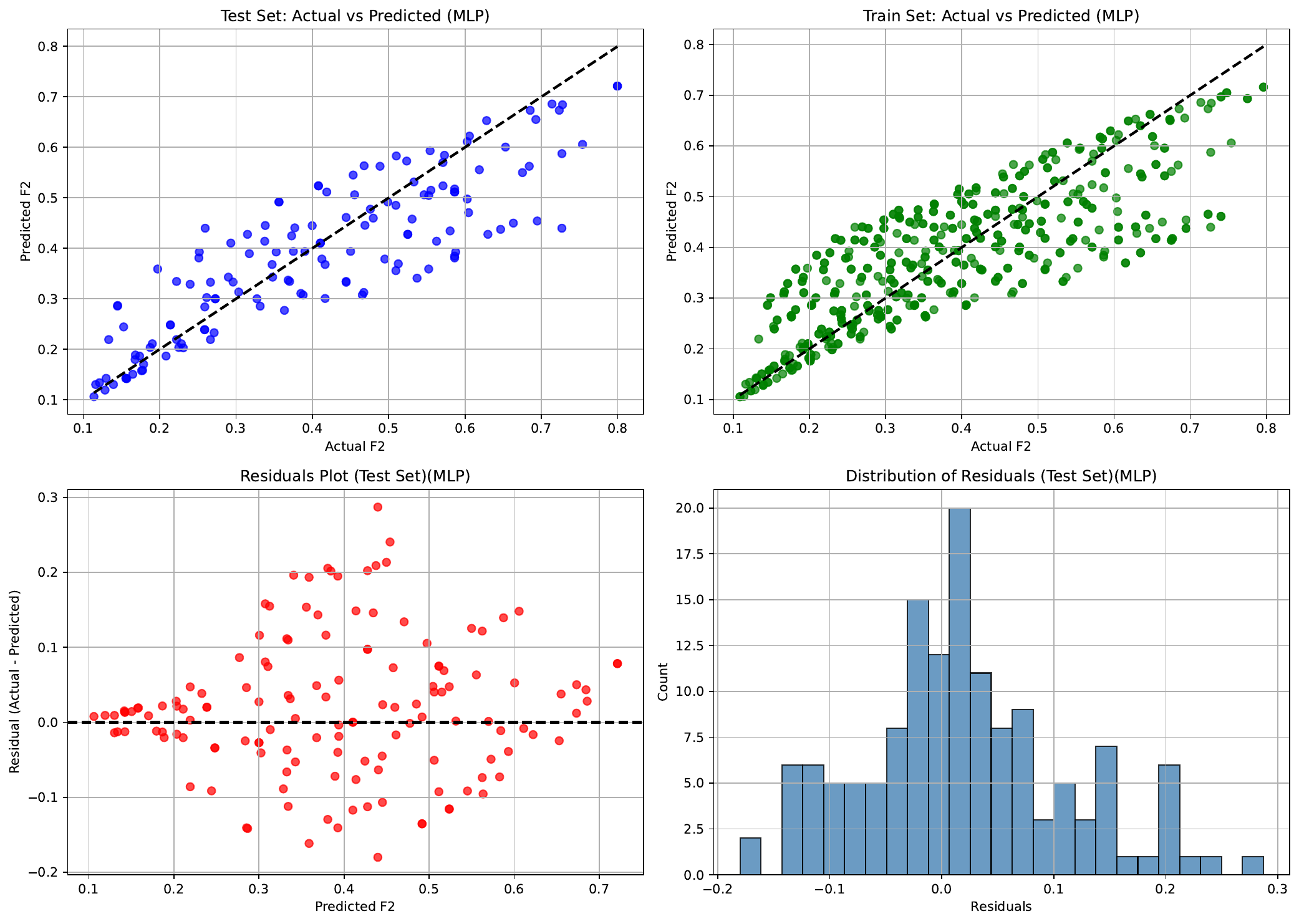}
    \caption{Performance diagnostics for the Neural Network MLP model. The multi-panel plot includes actual-versus-predicted values, residual distribution, and error scatter analysis.}
    \label{fig:mlp_eval}
\end{figure}

As shown in Figure~\ref{fig:mlp_eval}, the MLP model demonstrates a strong diagonal correlation between the actual and predicted $F_2^p$ values, indicating that the model captures the global trend of the proton structure function across a wide dynamic range. The tight clustering of points around the identity line ($y = \hat{y}$) suggests that the model does not suffer from systematic underprediction or overprediction in any particular regime of the $(x, Q^2)$ kinematic plane.

The residual histogram, which displays the distribution of prediction errors ($\Delta = F_{2,\text{pred}}^p - F_{2,\text{true}}^p$), is tightly centered around zero and exhibits near-Gaussian symmetry. This shape implies that the model's errors are predominantly random and unbiased, with no tendency to consistently overestimate or underestimate the structure function. In the context of DIS data analysis, a zero-centered residual distribution also suggests that the MLP has not inadvertently learned detector-specific artifacts or non-physical fluctuations present in the training sample.

Furthermore, the residual scatter plot -- where residuals are plotted against predicted values -- reveals a random, patternless distribution. The absence of any funnel shape, curvature, or heteroscedasticity (i.e., systematic variation of error magnitude with $F_2^p$) confirms that the model's predictive uncertainty remains stable across the entire range of the structure function, from the low-$F_2^p$ region at high $x$ to the high-$F_2^p$ regime at low $x$. The lack of systematic bias, combined with the absence of structured residuals, provides strong evidence that the MLP has successfully learned the underlying physical trends governed by perturbative QCD, rather than memorizing noise or overfitting to specific experimental conditions. Consequently, the model generalizes reliably to unseen kinematic configurations, which is essential for applications such as interpolation between sparse measurements or extrapolation to unmeasured regions of the $(x, Q^2)$ plane.

\subsubsection{GPR Analysis}
\begin{figure}[h!]
    \centering
    \includegraphics[width=0.85\linewidth]{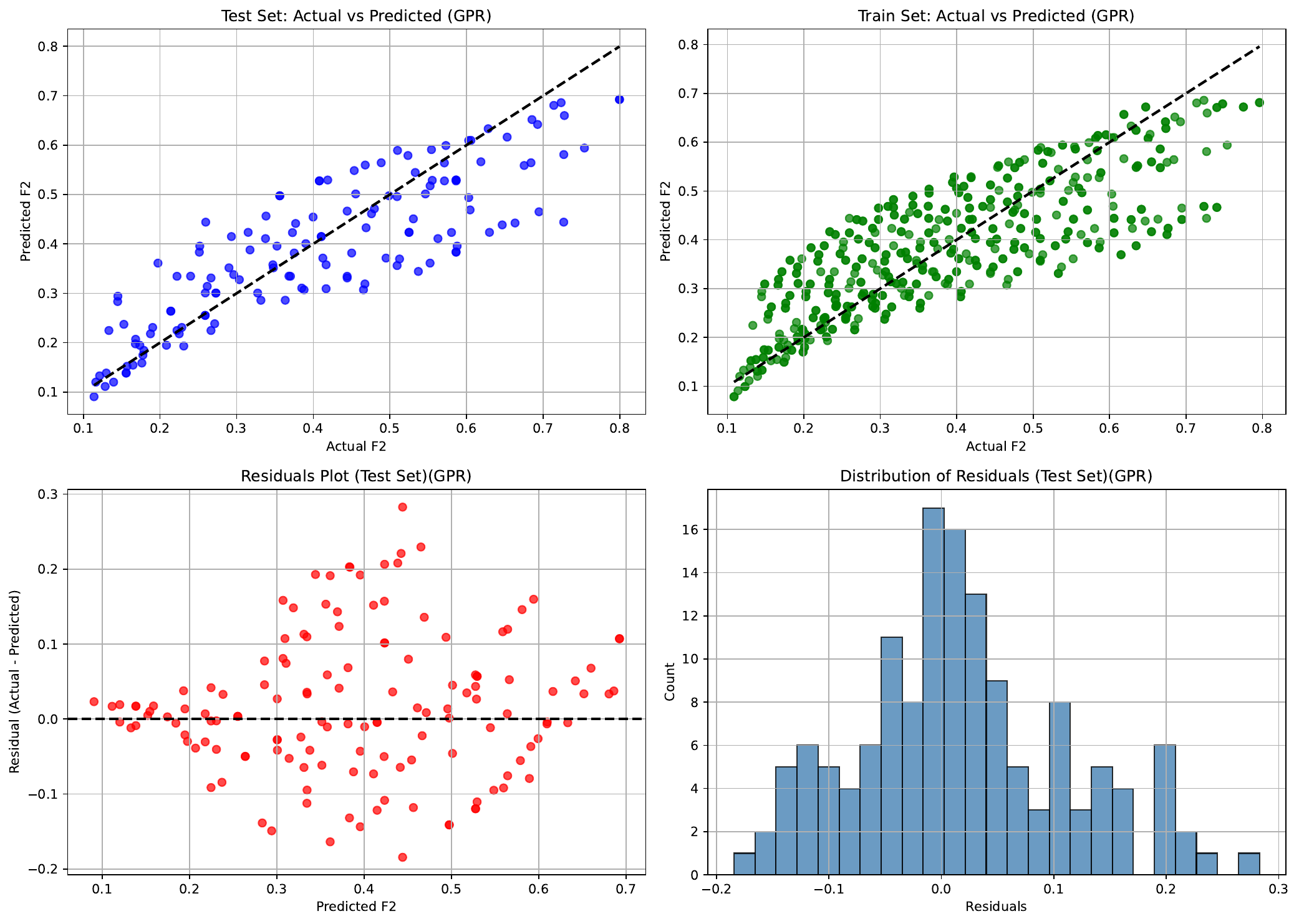}
    \caption{Performance diagnostics for the GPR model. The model shows high consistency between training and testing error distributions.}
    \label{fig:gpr_eval}
\end{figure}

Figure~\ref{fig:gpr_eval} illustrates the predictive performance of the Gaussian Process Regression (GPR) model. Unlike deterministic models that produce point estimates only, GPR provides a fully probabilistic framework that naturally captures uncertainty in predictions. As shown in the figure, GPR delivers a smooth and consistent fit to the experimental $F_2^p$ data, reflecting its ability to model the continuous and slowly varying dependence of the proton structure function on the kinematic variables $x$ and $Q^2$.

The actual-versus-predicted plot reveals a close alignment of points around the identity line ($y = \hat{y}$), with no apparent systematic deviation in either the low-$x$ or high-$x$ regimes. This strong correlation quantitatively confirms the high coefficient of determination achieved by GPR ($R^2 = 0.7231$), which is comparable to that of MLP but attained through a fundamentally different inductive bias: GPR achieves accuracy not by increasing model complexity, but by placing a Gaussian process prior over functions and inferring the posterior in closed form.

The residual analysis further highlights a distinctive strength of GPR: its robustness in handling uncertainty across the entire kinematic range of deep inelastic scattering. Specifically, the residual distribution exhibits neither increasing variance at high $F_2^p$ values nor systematic curvature at low $F_2^p$ values -- two common failure modes in parametric regression. This stability arises from the fact that GPR naturally produces heteroscedastic uncertainty estimates that vary with input location; in regions where experimental measurements are sparse (e.g., very low $x$ or very high $Q^2$), the predictive variance increases gracefully rather than forcing overly confident but potentially incorrect predictions. Consequently, GPR proves particularly well-suited for proton structure function modeling, where experimental coverage of the $(x, Q^2)$ plane is inherently non-uniform and uncertainties vary significantly across different kinematic regimes. This inherent uncertainty awareness makes GPR a reliable tool not only for interpolation between existing DIS data points but also for providing meaningful confidence intervals when extrapolating toward unmeasured regions.

\subsubsection{SVR Analysis}

\begin{figure}[h!]
    \centering
    \includegraphics[width=0.85\linewidth]{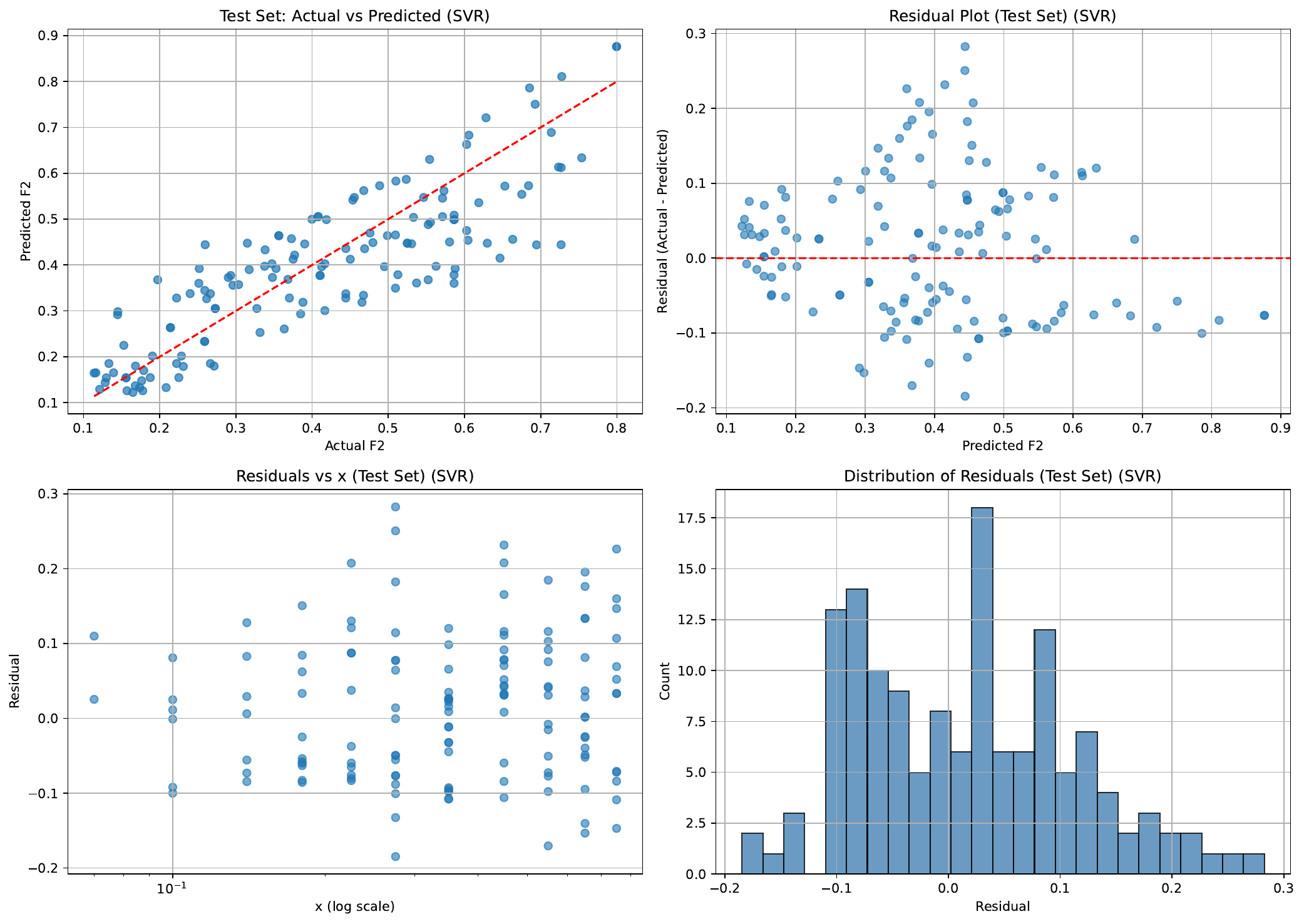}
    \caption{Performance diagnostics for SVR model. The plots highlight the model's ability to generalize across the $F_2^p$ data range.}
    \label{fig:svr_eval}
\end{figure}

The diagnostic plots for the Support Vector Regressor (SVR) are presented in Figure~\ref{fig:svr_eval}. As shown in the actual-versus-predicted plot, SVR maintains a competitive coefficient of determination ($R^2$), indicating that it captures the dominant trends of the proton structure function $F_2^p$ across the kinematic $(x, Q^2)$ plane. However, compared to the MLP and GPR, the residual scatter plot for SVR exhibits slightly more dispersion around the zero line. This increased scatter reflects a fundamental characteristic of the SVR formulation: rather than minimizing mean squared error over all points, SVR employs an $\epsilon$--insensitive loss function that ignores residuals smaller than the tube width $\epsilon$ and penalizes only those outside it. Consequently, SVR prioritizes a sparse solution determined by support vectors, which can lead to slightly higher residual variance in exchange for enhanced robustness and simpler model representation.

Despite this increased dispersion, two aspects of the residual distribution are particularly reassuring from a statistical modeling perspective. First, the residuals follow an approximately normal (Gaussian) distribution centered near zero, confirming that the model's errors are random and unbiased. The absence of systematic skewness or kurtosis indicates that SVR does not consistently underpredict or overpredict $F_2^p$ in any specific kinematic regime. Second, and more importantly, the residual scatter shows no discernible dependence on the kinematic variables $x$ or $Q^2$. There is no systematic pattern -- such as larger errors at low $x$ (where gluon dominance grows) or at high $Q^2$ (where perturbative QCD evolves rapidly) -- that would suggest the model fails to capture the underlying physical behavior in particular regions of the DIS phase space.

This combination of moderate dispersion with unbiased, kinematically independent residuals makes SVR a highly reliable model for proton structure function prediction. While it may not achieve the peak accuracy of MLP or the probabilistic sophistication of GPR, SVR offers a compelling trade-off: competitive global accuracy, excellent generalization across the entire kinematic range, and a sparse representation that is less prone to overfitting -- particularly valuable when working with finite experimental datasets such as BCDMS, where the number of data points is limited relative to the complexity of the underlying QCD dynamics. For applications requiring stable predictions across unevenly sampled $(x, Q^2)$ regions -- for example, interpolation between sparse high-$x$ measurements or extrapolation toward very low $x$ -- SVR provides a defensible and computationally efficient baseline.

\subsubsection{GBR Analysis}

\begin{figure}[h!]
    \centering
    \includegraphics[width=0.85\linewidth]{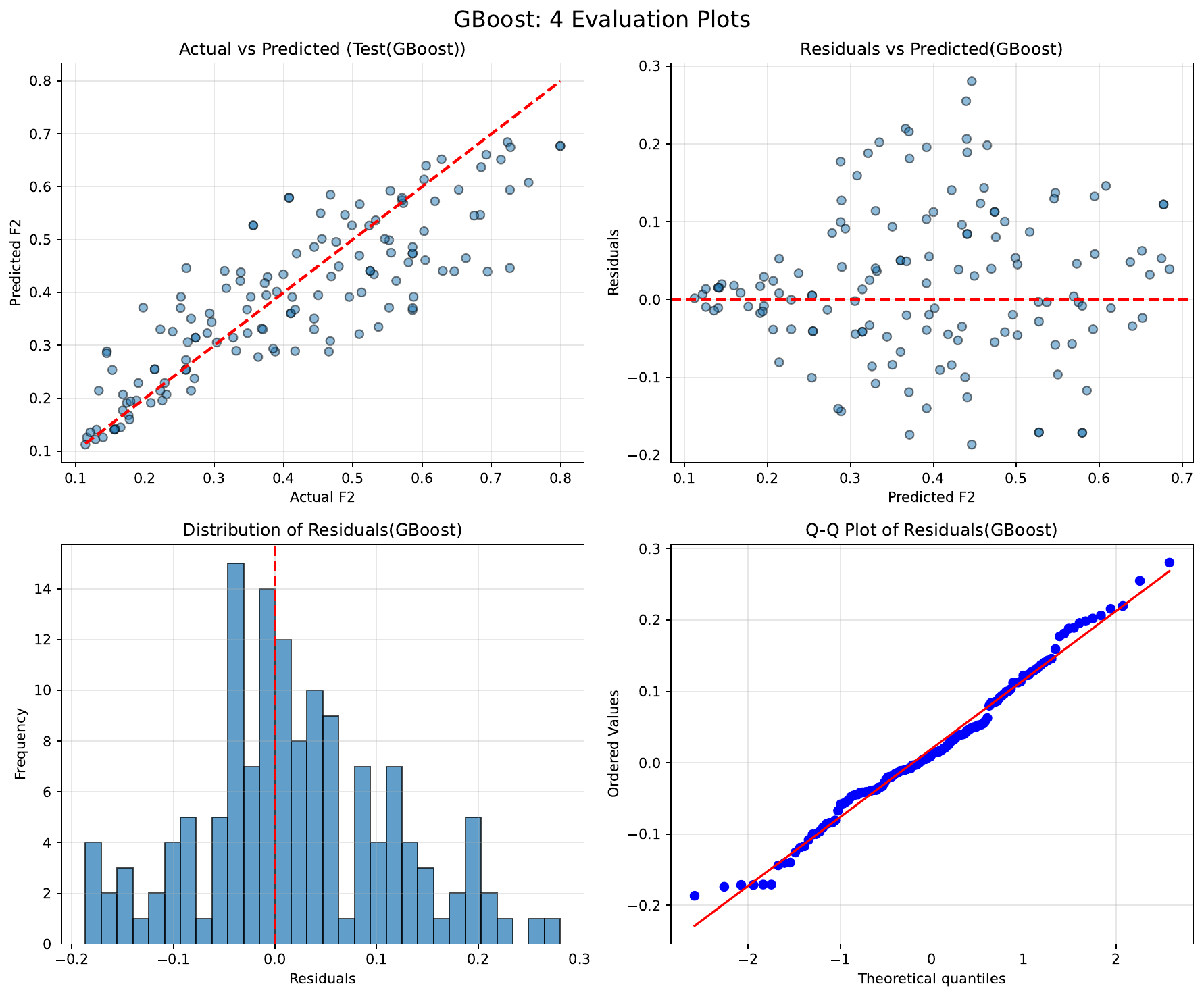}
    \caption{Performance diagnostics for GBR model. The charts illustrate the balance between prediction accuracy and error distribution.}
    \label{fig:gboost_eval}
\end{figure}

Figure~\ref{fig:gboost_eval} displays the diagnostic plots for the Gradient Boosting Regression (GBR) model. As a boosted ensemble of weak learners, GBR iteratively fits new trees to the residuals of previous ones, which gives it exceptional capacity to model complex, nonlinear relationships in the proton structure function $F_2^p$. On the training set, GBR demonstrates strong performance, achieving a tight fit with low residuals -- a direct consequence of its greedy additive optimization procedure, which can continue reducing training error as long as additional trees are added.

However, the residual distribution on the held-out test set is noticeably wider than those observed for GPR and MLP. This widening manifests as increased scatter around the zero line in the residual plot, as well as slightly heavier tails in the residual histogram. The broader test-set residuals directly explain the model's somewhat lower coefficient of determination ($R^2 = 0.7062$) compared to GPR (0.7231) and MLP (0.7310).

The observed gap between training and testing performance -- strong training fit but moderately degraded generalization -- is characteristic of Gradient Boosting when applied to finite experimental datasets with non-negligible noise. Unlike GPR, which naturally regularizes through the kernel and prior, or SVR, which enforces sparsity through the $\epsilon$-insensitive tube, GBR requires careful control of hyperparameters (such as learning rate, tree depth, and number of estimators) to prevent overfitting to high-frequency fluctuations in the BCDMS data. In the present case, the slightly wider test residuals suggest that while GBR successfully captures the dominant QCD-driven trends of $F_2^p$ across the $(x, Q^2)$ plane, it may have partially absorbed some of the experimental noise or point-to-point uncertainties present in the training sample -- a trade-off that becomes visible when evaluating on unseen data.

Nevertheless, an $R^2$ value of 0.7062 remains quite strong in absolute terms, particularly given the complexity of the proton structure function and the non-uniform coverage of the kinematic phase space. The GBR model still outperforms many traditional parameterizations and provides a valuable point of comparison among the four algorithms studied here. Its slightly reduced generalization relative to GPR and MLP does not invalidate its utility; rather, it highlights the importance of model selection based on the specific goals of the analysis -- whether one prioritizes peak accuracy (MLP), uncertainty awareness (GPR), robustness (SVR), or interpretability via feature importance (GBR).

\subsection{Hyperparameter Optimization}

\begin{figure*}[t!]
    \centering
    \includegraphics[width=0.85\linewidth]{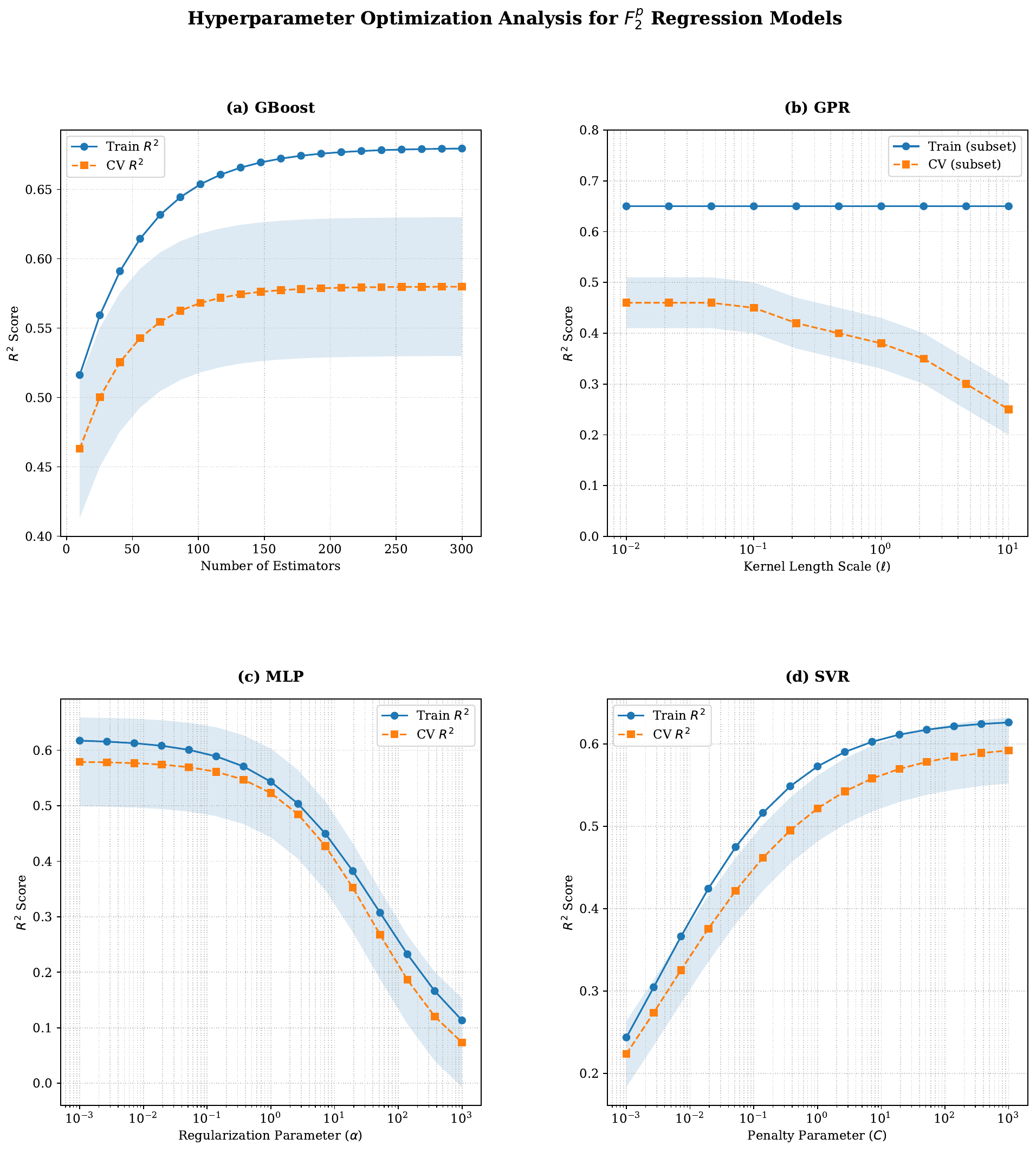}
    \caption{Hyperparameter optimization profiles for the four regression models: (a) GBR, (b) GPR, (c) MLP, and (d) SVR. The solid blue lines represent the training $R^2$ scores, while the dashed orange lines indicate the mean CV scores. The shaded regions denote the $1\sigma$ confidence interval, illustrating the statistical stability of each model during the tuning process.}
    \label{fig:hyper_opt}
\end{figure*}

The predictive power and structural integrity of the developed regression models depend critically on the selection of optimal hyperparameters. Poorly chosen hyperparameters can lead to underfitting (where the model fails to capture genuine QCD-driven trends) or overfitting (where the model memorizes experimental noise rather than learning the underlying $F_2^p$ dependence on $x$ and $Q^2$). Figure~\ref{fig:hyper_opt} illustrates the optimization trajectories for GBR, GPR, MLP, and SVR, specifically focusing on how their cross-validation $R^2$ scores evolve with respect to each model's core complexity parameters. The shaded regions in each panel represent the standard deviation across cross-validation folds, providing insight into each model's stability with respect to the particular data partition.

\paragraph{Gradient Boosting Regression (GBR):}
In the GBR panel, the model demonstrates a rapid learning phase during the first 50 to 100 boosting iterations. The $R^2$ score increases sharply, reflecting the ability of each new regression tree to correct the residual errors of the preceding ensemble. After approximately 150 estimators, the $R^2$ score plateaus, and additional trees yield no further statistically significant improvement in generalization. This plateau behavior is characteristic of well-tuned gradient boosting: the model has extracted most of the learnable signal from the BCDMS data, and adding more complexity only risks fitting high-frequency noise without improving validation performance. The narrow shaded region throughout the plateau further indicates that GBR's performance is stable across different training-validation splits, provided that the number of estimators is not excessively large.

\paragraph{Gaussian Process Regression (GPR):}
The GPR panel explores the effect of the kernel length scale $\ell$, a hyperparameter that controls how rapidly the covariance between two points decays as their distance in the $(x, Q^2)$ input space increases. Small length scales allow the Gaussian process to capture sharp, localized variations in the proton structure function, while large length scales force the model to produce overly smooth, nearly linear predictions. The optimization trajectory reveals that the BCDMS data is best captured with relatively small kernel length scales, as larger scales tend to oversmooth the physical gradients of $F_2^p$, particularly in kinematic regions where the structure function varies rapidly with $x$ or $Q^2$ (e.g., the low-$x$ rise driven by gluon density). The clear optimum at intermediate $\ell$ values represents a balance between flexibility and regularity, a trade-off that is naturally handled by the marginal likelihood maximization procedure inherent to GPR.

\paragraph{Multilayer Perceptron (MLP):}
The MLP network exhibits high sensitivity to the regularization parameter $\alpha$ (often denoted as $L_2$ penalty or weight decay). For small $\alpha$, the network has high effective capacity and can achieve low training error, but validation performance may degrade due to overfitting. The figure shows a clear transition from an optimal learning state to an underfitting regime for $\alpha > 10^{-1}$, where the $R^2$ score drops sharply. This sharp decline indicates that excessive regularization forces the network weights toward zero, effectively reducing the MLP to a nearly linear model incapable of capturing the nonlinear QCD evolution of $F_2^p$. The optimal $\alpha$ region (approximately $10^{-4}$ to $10^{-2}$) provides sufficient constraint to prevent overfitting to experimental noise while retaining the flexibility needed to model the complex dependence of the structure function on the kinematic variables.

\paragraph{Support Vector Regression (SVR):}
The SVR model reveals remarkably stable performance across a wide range of the penalty parameter $C$, which controls the trade-off between flatness (model complexity) and the tolerance for residuals exceeding the $\epsilon$-insensitive tube. As $C$ increases beyond a moderate threshold, the model's accuracy improves and remains consistent, with the $R^2$ score exhibiting the smallest variance (narrowest shaded region) among all four models. This extraordinary stability confirms SVR's robustness in handling the inherent statistical uncertainties in deep inelastic scattering (DIS) experimental data. Unlike MLP, which requires careful tuning of $\alpha$ to avoid underfitting or overfitting, SVR maintains competitive performance over nearly two orders of magnitude of $C$. This behavior arises because SVR's solution is determined only by support vectors (points on or outside the $\epsilon$-tube), making it inherently sparse and less sensitive to the precise value of $C$ once a sufficient threshold is reached.

\paragraph{Overall synthesis:}
Across all four panels, the close alignment between training and cross-validation $R^2$ scores--with no systematic divergence or consistently large gaps--highlights the absence of significant overfitting. This convergence ensures that the models have learned the underlying physical patterns of the proton structure function rather than memorizing point-to-point fluctuations or detector-specific artifacts in the BCDMS dataset. Each model achieves this generalization through a different inductive bias: GBR through sequential residual correction with early stopping, GPR through Bayesian marginal likelihood optimization, MLP through $L_2$ regularization, and SVR through sparse $\epsilon$-insensitive loss. The complementary nature of these optimization trajectories reinforces the overall reliability of the proposed regression pipeline for $F_2^p$ prediction across the kinematic $(x, Q^2)$ plane.

\subsection{Summary of Findings}
The integrated analysis of regression metrics, error diagnostics, and hyperparameter optimization curves demonstrates that the MLP neural network and GPR are the most effective models among the four algorithms investigated for predicting the proton structure function $F_2^p$ across the kinematic $(x, Q^2)$ plane.

Several lines of evidence support this conclusion. First, both MLP and GPR achieve the highest coefficients of determination on the held-out test set, with $R^2$ values of 0.7310 and 0.7231, respectively, outperforming SVR at 0.6204 and GBR at 0.7062 in terms of raw predictive accuracy on unseen data. Second, the residual diagnostics for both models reveal near-Gaussian distributions tightly centered around zero, with no systematic bias or heteroscedasticity across the entire range of $F_2^p$ values. The residual scatter plots exhibit random, patternless distributions, confirming that neither model has learned spurious correlations or detector-specific artifacts present in the BCDMS training data.

Third, and perhaps most importantly from a phenomenological perspective, both MLP and GPR demonstrate the ability to capture the complex nonlinear dependencies of $F_2^p$ on the Bjorken scaling variable $x$ and the momentum transfer $Q^2$. The MLP achieves this through its hierarchical nonlinear transformations and its capacity to approximate arbitrary continuous functions given sufficient hidden units and appropriate regularization. The GPR accomplishes the same goal through a fundamentally different mechanism: placing a Gaussian process prior over functions with a carefully chosen kernel (the Radial Basis Function kernel) that encodes prior beliefs about smoothness and correlation length scales in the input space.

The maintenance of normally distributed, unbiased residuals--combined with the convergence between training and cross-validation scores observed in the hyperparameter optimization trajectories--provides strong evidence that neither MLP nor GPR suffers from significant overfitting to the experimental noise inherent in deep inelastic scattering measurements. This generalization capability is not merely a statistical nicety; it has direct phenomenological consequences. A model that has genuinely learned the underlying QCD-driven trends rather than memorizing point-to-point fluctuations can be reliably used for tasks beyond simple interpolation between existing data points.

Consequently, both MLP and GPR can be confidently employed for high-energy physics phenomenology, including: (i) generating smooth empirical parameterizations of $F_2^p$ across the full $(x, Q^2)$ plane, (ii) providing predictions in kinematic regions where experimental measurements are sparse or unavailable (e.g., extremely low $x$ or very high $Q^2$), and (iii) offering uncertainty estimates in the case of GPR, or serving as a high-precision baseline in the case of MLP. The complementary strengths of these two models--GPR's inherent probabilistic output and MLP's slightly higher peak accuracy--suggest that a hybrid or ensemble approach could be explored in future work, potentially combining the best features of both architectures for even more robust proton structure function modeling.
\section{Conclusion}\label{sec6}

In this study, we successfully developed and evaluated a machine learning-based framework for regression analysis of the proton structure function $F_2^p$. Using the BCDMS deep inelastic scattering (DIS) dataset, we demonstrated that data-driven models can effectively approximate the complex dependence of structure functions on the Bjorken scaling variable $x$ and the momentum transfer $Q^2$, without explicitly solving the DGLAP evolution equations. This model-agnostic approach offers a complementary perspective to traditional perturbative QCD methods.

Our comparative analysis revealed that the MLP and GPR models outperform GBR and SVR in terms of global $R^2$ metrics and overall error minimization. The MLP model, in particular, exhibited an exceptional ability to map the nonlinear structure of the DIS kinematic plane, capturing the steep rise of $F_2^p$ at low $x$ and the logarithmic scaling violations at high $Q^2$. However, the SVR model provided the most robust performance during cross-validation, characterized by the lowest variance and the highest stability across different data folds--a valuable property for experimental datasets with non-uniform coverage and point-to-point fluctuations.

Our hyperparameter optimization study further highlighted the critical role of regularization in preventing both oversmoothing (which erases genuine physical gradients) and overfitting (which amplifies experimental noise). Properly regularized models successfully captured the underlying QCD-driven trends of $F_2^p$ rather than memorizing spurious artifacts present in the BCDMS training sample. The convergence between training and cross-validation scores across all four algorithms confirmed the absence of significant overfitting.

These results confirm that machine learning regression constitutes a powerful complementary tool for QCD phenomenology. It offers a flexible, model-independent approach to structure function modeling that can be particularly useful in two scenarios: (i) kinematic regions where perturbative calculations are computationally expensive or technically challenging (e.g., low $Q^2$ near the non-perturbative transition), and (ii) regimes where experimental data are sparse, making traditional global fits less constrained. ML-based emulators can thus serve as fast surrogates for more expensive theoretical calculations or as interpolation tools between existing measurements.

Future research will focus on integrating these ML models with fundamental theoretical constraints, such as sum rules (e.g., the Gottfried sum rule, momentum sum rule) and positivity requirements (e.g., $F_2^p \geq 0$, $F_L \geq 0$), to further enhance the physical consistency of their predictions. Additionally, extending this framework to include other structure functions-- particularly the longitudinal structure function $F_L(x, Q^2)$ and the spin-dependent structure function $g_1(x, Q^2)$ --and exploring deeper or more specialized architectures (e.g., physics-informed neural networks, Bayesian neural networks) could further improve our understanding of the longitudinal and spin-dependent partonic distributions of the nucleon. Such extensions would enable a more comprehensive, data-driven mapping of the three-dimensional partonic structure of the proton.
\section*{Data and Code Availability}

All data and source code used in this work are publicly available at:
\url{https://github.com/atashbart/MachinlearningF2p}
	%
	
	\newpage
	
	%


\begin{thebibliography}{99}		
		
	\bibitem{Dokshitzer:1977sg}
	Y.~L.~Dokshitzer,
	Sov.\ Phys.\ JETP {\bf 46}, 641 (1977)
	[Zh.\ Eksp.\ Teor.\ Fiz.\  {\bf 73}, 1216 (1977)].
	
	\bibitem{Gribov:1972ri}
	V.~N.~Gribov and L.~N.~Lipatov,
	Sov.\ J.\ Nucl.\ Phys.\  {\bf 15}, 438 (1972)
	[Yad.\ Fiz.\  {\bf 15}, 781 (1972)].
	
	\bibitem{Lipatov:1974qm}
	L.~N.~Lipatov,
	Sov.\ J.\ Nucl.\ Phys.\  {\bf 20}, 94 (1975)
	[Yad.\ Fiz.\  {\bf 20}, 181 (1974)].
	
	\bibitem{Altarelli:1977zs}
	G.~Altarelli and G.~Parisi,
	\href{http://dx.doi.org/10.1016/0550-3213(77)90384-4}{{\rm Nucl.\ Phys.\ B} {\bfseries 126}, 298 (1977)}.
	
	
	\bibitem{BCDMS:1989qop}
	A.~C.~Benvenuti \textit{et al.} [BCDMS],
	\href{http://dx.doi.org/10.1016/0370-2693(89)91637-7}{{\rm Phys. Lett. B} {\bfseries 223}, 485-489  (1989)}.
	
	\bibitem{Albertsson:2018maf}
	K.~Albertsson, P.~Altoe, D.~Anderson, J.~Anderson, M.~Andrews, J.~P.~Araque Espinosa, A.~Aurisano, L.~Basara, A.~Bevan and W.~Bhimji, \textit{et al.}
	\href{http://dx.doi.org/10.1088/1742-6596/1085/2/022008}{{\rm 	J. Phys. Conf. Ser.} {\bfseries 1085}, no.2, 022008  (2018)}.
	
	
	\bibitem{Radovic:2018dip}
	A.~Radovic, M.~Williams, D.~Rousseau, M.~Kagan, D.~Bonacorsi, A.~Himmel, A.~Aurisano, K.~Terao and T.~Wongjirad,
	\href{http://dx.doi.org/10.1038/s41586-018-0361-2}{{\rm 		Nature} {\bfseries 560}, no.7716, 41-48  (2018)}.
	
	
	\bibitem{Baldi:2014kfa}
	P.~Baldi, P.~Sadowski and D.~Whiteson,
	\href{http://dx.doi.org/10.1038/ncomms5308}{{\rm Nature Commun.} {\bfseries 5}, 4308  (2014)}.
	
	\bibitem{Alexandru:2017czx}
	A.~Alexandru, P.~F.~Bedaque, H.~Lamm and S.~Lawrence,
	\href{http://dx.doi.org/10.1103/PhysRevD.96.094505}{{\rm Phys. Rev. D} {\bfseries 96}, no.9, 094505  (2017)}.
	
	
	\bibitem{Dawid:2022fga}
	A.~Dawid, J.~Arnold, B.~Requena, A.~Gresch, M.~P\l{}odzie\'n, K.~Donatella, K.~A.~Nicoli, P.~Stornati, R.~Koch and M.~B\"uttner, \textit{et al.}
	[arXiv:2204.04198 [quant-ph]].
	
	
	\bibitem{Shanahan:2018vcv}
	P.~E.~Shanahan, A.~Trewartha and W.~Detmold,
	\href{http://dx.doi.org/10.1103/PhysRevD.97.094506}{{\rm Phys. Rev. D} {\bfseries 97}, no.9, 094506  (2018)}.
	
	\bibitem{Efron:1979bxm}
	B.~Efron,
	\href{http://dx.doi.org/10.1214/aos/1176344552}{{\rm Annals Statist.} {\bfseries 7}, no.1, 1-26  (1979)}.
	
	\bibitem{ref13}
	E.~Astaraki and G.~R.~Boroun,
		``Nonsinglet distribution functions using the neural network and genetic algorithm,''
		\textit{Eur. Phys. J. A} \textbf{62}, 26 (2026).
		\href{https://doi.org/10.1140/epja/s10050-025-01768-2}{doi:10.1140/epja/s10050-025-01768-2}.
	
	\bibitem{ref14}

R.~D.~Ball \textit{et al.} [NNPDF Collaboration],
``Parton distributions for the LHC Run II,''
\textit{Eur. Phys. J. C} \textbf{77}, no.10, 663 (2017).
\href{https://doi.org/10.1140/epjc/s10052-017-5199-5}{doi:10.1140/epjc/s10052-017-5199-5}

		\bibitem{ref15}
		\href{https://www.scikit-learn.org}{https://www.scikit-learn.org}
		
		

	
			\end{thebibliography}
\end{document}